\documentclass[structabstract]{aa}  
\usepackage{graphicx}
\usepackage{txfonts}
\usepackage{natbib}
\def\igr{IGR~J09026$-$4812}
\def\integral{\emph{INTEGRAL}}

\def\chandra{\emph{Chandra}}

\def\cps{counts$\,\mathrm{s}^{-1}$}
\newcommand{\unit}[2]{\mathrm{#1}^{#2}}
\def\ecms{\mathrm{erg}\,\unit{cm}{-2}\,\unit{s}{-1}}
\def\es{\mathrm{ergs}\,\unit{s}{-1}}

\def\mum{\mu\mathrm{m}}
\def\Ks{$K_{\mathrm{S}}$}

\def\nh{N_{\mathrm{H}}}

\newcommand\ra[3]{#1^{\mathrm{h}}#2^{\mathrm{m}}#3^{\mathrm{s}}}
\newcommand\dec[3]{#1\degr#2\arcmin#3\arcsec}
\begin{document}
   \title{Infrared identification of \igr\ as a Seyfert 1 galaxy\thanks{Based on observations made with ESO Telescopes at the La Silla Observatory under programme ID 078.D$-$0268(B).}}

   \author{J.A.~Zurita Heras
          \inst{1}
          \and
          S.~Chaty\inst{1}
          \and
          J.~A. Tomsick\inst{2}
          }

   \institute{Laboratoire AIM, CEA/DSM-CNRS-Universit\'e Paris Diderot,
	IRFU/Service d'Astrophysique, 91191 Gif-sur-Yvette, France\\
	\email{juan-antonio.zurita-heras@cea.fr}; \email{sylvain.chaty@cea.fr}
	\and
	Space Sciences Laboratory, 7 Gauss Way, University of California, Berkeley, CA 94720$-$7450, USA\\
	\email{jtomsick@ssl.berkeley.edu}
   }

   \date{Received --; accepted --}

 
  \abstract
   {\igr\ was discovered by \integral\ in 2006 as a new hard X-ray source. Thereafter, an observation with \chandra\ pinpointed a single X-ray source within the ISGRI error circle, showing a hard spectrum, and improving its high-energy localisation to a subarcsecond accuracy. Thus, the X-ray source was associated to the infrared counterpart 2MASS~J09023731$-$4813339 whose $JHK_{\mathrm{S}}$ photometry indicated a highly reddened source. The high-energy properties and the counterpart photometry suggested a high-mass X-ray binary with a main sequence companion star located 6.3--8.1~kpc away and with a 0.3--10~keV luminosity of $8_{-1}^{+13}\times10^{34}\,\es$.}
   {New optical and infrared observations were needed to confirm the counterpart and to reveal the nature of \igr.}
   {We performed optical and near infrared observations on the counterpart 2MASS~J09023731$-$4813339 with the ESO/NTT telescope on March 2007. We achieved photometry and spectroscopy in near infrared wavelengths and photometry in optical wavelengths.}
   {The accurate astrometry at both optical and near infrared wavelengths confirmed 2MASS~J09023731$-$4813339 to be the counterpart of \igr. However, the near infrared images show that the source is extended, thus excluding any Galactic compact source possibility. The source spectrum shows three main emission lines identified as the HeI $\lambda\,1.0830\,\mum$ line, and the HI Pa$\,\beta$ and Pa$\,\alpha$ lines, typical in galaxies with an active galactic nucleus. The broadness of these lines reached values as large as 4000~km$\,\unit{s}{-1}$ pointing towards a type 1 Seyfert galaxy. The redshift of the source is $z=0.0391\pm0.0004$. Thus, the near infrared photometry and spectroscopy allowed us to classify \igr\ as a Seyfert galaxy of type 1.}
   {}


   \maketitle
%

\section{Introduction}

The ESA high-energy space mission \integral\ \citep{Winkleral03} has performed a deep survey of our Galaxy allowing the discovery of many new hard X-ray sources. This has been mainly achieved with the coded-mask imager IBIS/ISGRI \citep[15~keV--1~MeV, field of view of $29\degr\times29\degr$, angular resolution of 12$\arcmin$;][]{Ubertinial03,Lebrunal03}. The 20--100~keV hard X-ray emission as observed by ISGRI has been regularly reported \citep{Birdal06,Birdal07,Bodagheeal07}. The unclassified source \igr\ was first reported in the 2nd ISGRI catalogue as a new hard X-ray source \citep{Birdal06}. Its position was R.A.~(2000)~$=135\degr.638$ and Dec.~$=-48\degr.196$ ($1.1\arcmin$ at 68\% confidence level). The average 20--40 and 40--100 keV fluxes were $0.9\pm0.1$ and $0.9\pm0.2$~mCrab, respectively, for an effective exposure of 1.35~Ms.
These values only suffered a slight change in the 3rd ISGRI catalogue \citep{Birdal07}: R.A.~(2000)~$=135\degr.668$, Dec.~$=-48\degr.216$ ($2.3\arcmin$ at 90\% confidence level), $F_{\mathrm{20-40keV}}=1.3\pm0.1$~mCrab and $F_{\mathrm{40-100keV}}=1.4\pm0.2$~mCrab, for an effective exposure of 1.5~Ms.
The source \igr\ was tentatively associated with a faint \emph{ROSAT} counterpart located $2.5\arcmin$ away from the ISGRI position at R.A.~(2000)~$=\ra{09}{02}{38.4}$ and Dec.~=$\dec{-48}{14}{08.0}$ ($20\arcsec$) that showed an average count rate of $(2.7\pm0.9)\times10^{-2}$~\cps \citep{Stephenal06}.

\citet{Tomsickal08a} reported a 5~ks \chandra\ observation performed on Feb.~5, 2007, that showed a single X-ray source within the ISGRI error circle located at R.A.~(2000)~$=\ra{09}{02}{37.33}$ and Dec.~=$\dec{-48}{13}{34.1}$ ($0.6\arcsec$ at 90\% confidence level). This position is located outside the \emph{ROSAT} error circle, discarding the soft X-ray counterpart suggested by \citet{Stephenal06}. The 0.3--10~keV spectrum was fitted with an absorbed power law model: $\nh=(1.9_{-0.4}^{+0.6})\times 10^{22}\ \unit{cm}{-2}$, $\Gamma=1.1_{-0.3}^{+0.5}$ (90\% confidence errors) and an unabsorbed 0.3--10~keV flux of $1.3\times 10^{-11}\ \ecms$. They also showed that the source was not intrinsically absorbed comparing the fitted $\nh$ to the atomic and molecular hydrogen column densities through our Galaxy. They suggested the source 2MASS~J09023731$-$4813339 as the infrared (IR) counterpart since this source was located $0.4\arcsec$ away from the \chandra\ position. Its IR magnitudes were $J=15.57\pm0.08$, $H=13.86\pm0.07$, and $K_{\mathrm{S}}=12.69\pm0.04$. Due to its \Ks\ brightness, high extinction, and hard power law, \citet{Tomsickal08a} suggested that \igr\ was a new high-mass X-ray binary (HMXB) with a main sequence companion star.

We have performed optical and near infrared (NIR) observations on \igr\ in order to determine its nature. We describe the observations and the data reduction in Sect.~\ref{secObs}. We then present the results and discuss the nature of this source in Sect.~\ref{secDis}, and conclude in Sect.~\ref{secCon}.


\section{Observations and data analysis}\label{secObs}

\begin{figure*}
\centering
\resizebox{0.6\textwidth}{!}{\includegraphics{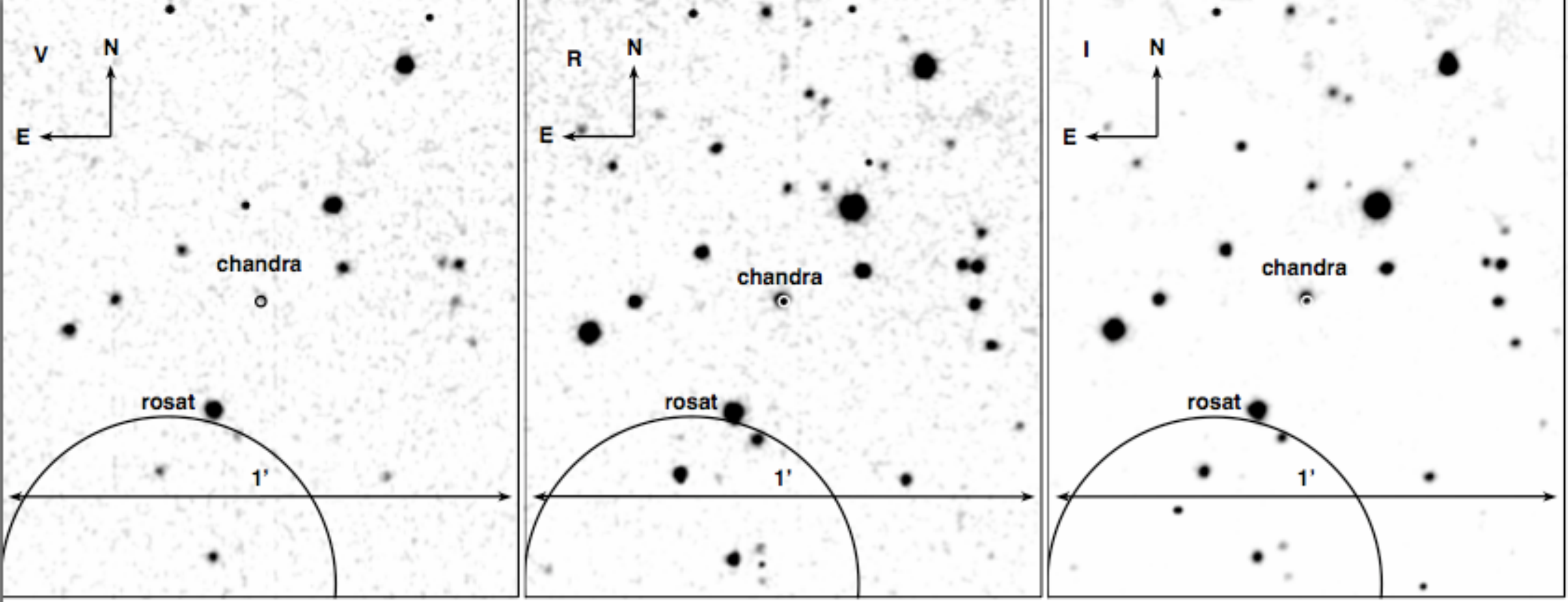}}
\resizebox{0.6\textwidth}{!}{\includegraphics{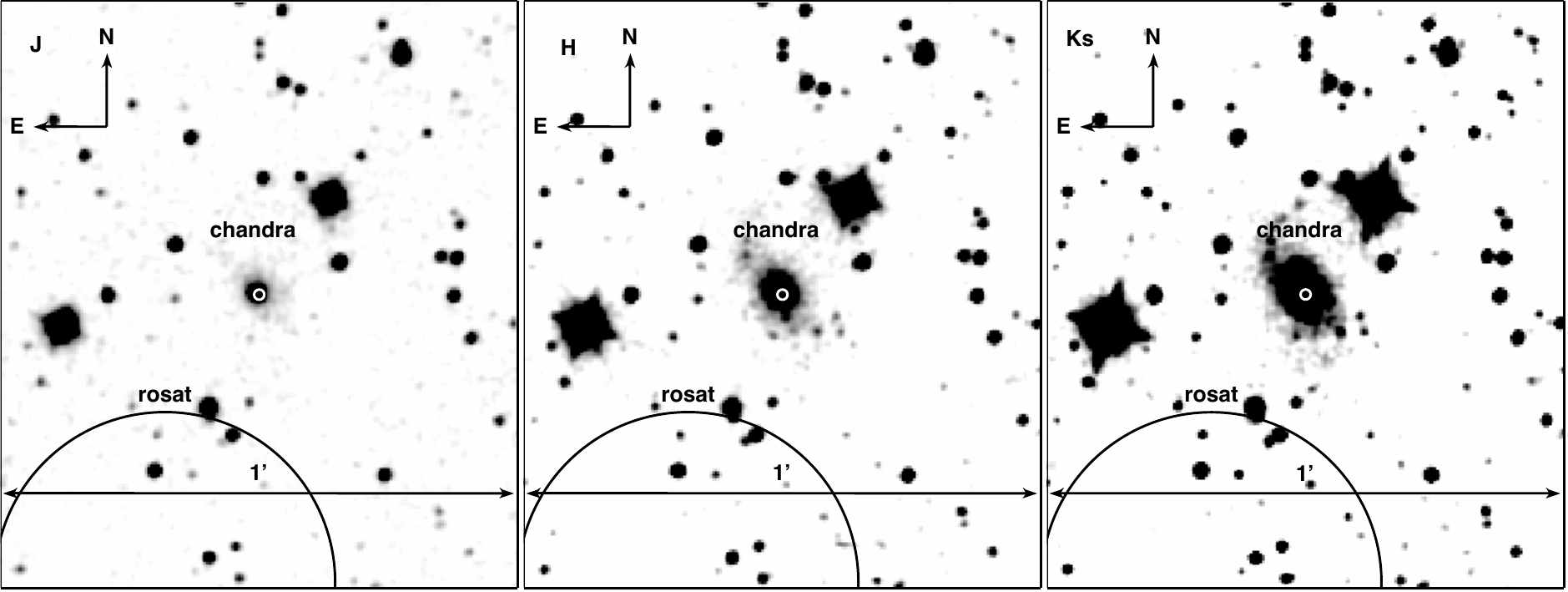}}
\caption{\emph{Left to right, top to bottom:} $VRI$ and $JHK_{\mathrm{s}}$ images taken on March 10 (NIR) and 14 (optical), 2007.}
\label{IRima}
\end{figure*}

The observations were carried out with the ESO 3.6~m New Technology Telescope (NTT) at La Silla Observatory, Chile, as part of the program 078.D--0268(B) through service mode. The optical and NIR observations were achieved with the imager SUSI--2 (350--900~nm) and the spectro-imager SOFI (0.9--2.4~$\mu$m), respectively. Both instruments are installed on the same Nasmyth focus of the NTT. Astrometry, photometry and spectroscopy were achieved during these observations.

\subsection{NIR observations}

NIR photometry in the bands $J$, $H$ and $K_{\mathrm{S}}$ were performed on March 5 and 10, 2007. The observations were centred on the accurate X-ray position of \igr. The large imaging mode was used during those observations that implied an image pixel scale of $0.288\arcsec$/pixel and a field of view of $4.92\arcmin\times 4.92\arcmin$. Nine images were taken for each filter $J$, $H$ and \Ks\ with integration times of 60~s each. For each filter, four of the nine images were taken with a slight offset of $\sim30\arcsec$ that allowed us to build the NIR sky in order to subtract it from the images. Two standard stars chosen in the faint NIR standard star catalogue \citep{Perssonal98} were also observed: S255$-$S and S262$-$E. Five observations per filter were performed on each standard star. The first observation was centred on the target, and then, the next four images were taken with an offset of $\sim45\arcsec$ compared to the first one. The same strategy with the same standard stars was applied during the 2nd night. The seeing conditions varied from $\sim0.9\arcsec$ during the 1st night to $\sim0.5\arcsec$ during the 2nd night.

NIR spectroscopy using both low-resolution Blue and Red grisms was carried out on March 5 and 12, 2007. For each filter, twelve spectra were taken with the $1.0\arcsec$ slit. Half of the spectra were taken on the source and the other half with an offset of $60\arcsec$ in order to subtract the NIR sky from the images, the jitter pattern being AAA$\,$BBB$\,$BBB$\,$AAA. The integration time was 180~s for each spectrum for a total observation of 4320~s. Two spectroscopic standard stars were observed: HD62388 (spectral type (sp.T.) A0V) and Hip0842847 (sp.T. G2V) on March 5 and 12, respectively. For each standard star, four spectra in both Blue and Red grisms were taken, of which half of them were taken with an offset of 45$\arcsec$.

\subsection{Optical observations}

Only optical photometry was carried out on March 14, 2007, with the filters $U,B,V,R,I,Z$. One image per filter was obtained with a field of view of $5.5\arcmin\times5.5\arcmin$ with a binning factor of 2 that implied a pixel scale of 0.161$\arcsec$/pixel. The integration time was 60~s in each filter. Thirteen photometric standard stars selected in the optical standard star catalogue of \citet{Landolt92} were observed in the 2 fields 98\_(733,1087,1102,1112,1119,1122,1124) and  RU\_152$+$RU\_152(A,B,C,E,F). The integration times varied between 5 and 30~s.

\subsection{Data reduction}

The reduction of both optical and NIR data was performed with the Image Reduction and Analysis Facility (IRAF\footnote{IRAF is available at {\tt http://iraf.net/}}) version 2.13beta2. Data reduction was performed using standard procedures on the optical and NIR images, including the crosstalk, the correction of the dark current (in optical), the flat-fielding and the NIR sky subtraction.

We performed accurate astrometry on each image ($U,B,V,R,I,Z,J,H,K_{\mathrm{S}}$) using the {\tt gaia} tool from the Starlink suite and using the 2MASS catalogue for the NIR images and the USNO B1.0 catalogue for the optical images. The NIR root mean square (rms) of the astrometrical fit was always lower than $0.4\arcsec$ with the expected pixel scale in x,y axis of $-0.288\times0.288\arcsec$/pixel. The optical rms of the astrometrical fit was also lower than $0.4\arcsec$ with the expected pixel scale in x,y axis of $0.161\times0.161\arcsec$/pixel. 

We carried out aperture photometry in a crowded field using the IRAF {\tt digiphot.daophot} package. For the NIR photometry, the point sources located near \igr\ were subtracted using the same package. Aperture photometry on the source was performed with several ellipses using the {\tt stsdas.isophot} and {\tt digiphot.apphot} packages. The instrumental magnitudes $m_{\mathrm{instr}}$ were transformed into apparent magnitudes $m_{\mathrm{app}}$ using the standard photometric relation: $m_{\mathrm{app}}=m_{\mathrm{instr}}-Z_{\mathrm{p}}-ext\times AM$, where $Z_{\mathrm{p}}$ is the zero-point, $ext$ the extinction and $AM$ the airmass. The colour term was not used in the NIR because there were not enough standard stars nor at optical wavelengths because we did not detect the sources of interest in several bands, particularly in $V$. The $Z_{\mathrm{p}}$ parameters were fitted using the photometric relation in order to match the instrumental and apparent magnitudes of the standard stars. The extinction parameters were fixed to the values given in the SOFI and SUSI-2 calibration tables\footnote{retrieved at {\tt http://www.eso.org/sci/facilities/lasilla/}\\ {\tt telescopes/ntt/index.html}}. The magnitudes, airmass, zero-point and extinction parameters are reported in Table~\ref{tab_photom}. Images in $VRIJHK_{\mathrm{s}}$ are shown in Fig.~\ref{IRima}. The NIR surface brightness profiles are shown in Fig.~\ref{brightima}.

The NIR spectra were reduced using the IRAF {\tt noao.twodspec} package. Each individual image was corrected using standard procedures for the crosstalk and the flat-fielding. Consecutive images with the same jitter were combined together to obtain 4 output images. They were corrected for the NIR sky, subtracting images with different jitter. We extracted the source spectrum from the 4 images. We performed the wavelength calibrations using a xenon arc extracted with the same parameters as for the source. The 4 spectra were combined in a single spectrum and corrected from telluric features using the spectrum of the standard star and the IRAF tool {\tt telluric}. We applied this strategy for both blue and red filters and we finally combined the blue and red spectra. We multiplied the full spectrum by a calibrated spectrum of an A$0$~V (or G$2$~V) star from the spectral library of \citet{Pickles98} to avoid spectral contamination from the telluric star. The output spectra were dereddened from the Galactic extinction using $E(\mathrm{B}-\mathrm{V})=1.509$ \citep{Schlegelal98}. The resulting spectra are shown in Fig.~\ref{SPEima}. 


\section{Results and discussion}\label{secDis}

Only one optical and NIR candidate was located within the \chandra\ error circle (see Fig.~\ref{IRima}). Our NIR astrometry confirmed the candidate 2MASS~J09023731$-$4813339 suggested by \citet{Tomsickal08a}. 

\begin{table}
\caption{NIR and optical photometry of the single counterpart.}
\label{tab_photom}      
\centering          
\begin{tabular}{c c c c}
\hline\hline       
Date & \multicolumn{3}{c}{2007--03--05T01:27--02:00}\\
Filters & $J$ & $H$ & \Ks \\
\hline
Mag.$^\diamond$ & $15.2 \pm0.1$ & $13.66\pm0.03$ & $12.3 \pm0.2$\\
$Z_{\mathrm{p}}$ & $2.0\pm0.1$ & $2.12\pm0.03$ & $2.9\pm0.2$\\
$AM^\dagger$ & 1.10 & 1.09 & 1.08\\
SMA$[\,\arcsec]$ & 3.3 & 4.0 & 4.9 \\
P.A.$[\,\degr]$ & $-50\pm11$ & $-60\pm3$ & $-51\pm2$\\
Ecc. & $0.51\pm0.09$ & $0.75\pm0.03$ & $0.78\pm0.02$\\ 
\hline\hline
Date & \multicolumn{3}{c}{2007--03--10T00:21--00:41}\\
Filters & $J$ & $H$ & \Ks \\
\hline
Mag.$^\diamond$ & $15.24\pm0.02$ & $13.63\pm0.02$ & $12.53\pm0.08$  \\
$Z_{\mathrm{p}}$ & $1.929\pm0.002$ & $2.07\pm0.02$ & $2.62\pm0.07$\\
$AM^\dagger$ & 1.20 & 1.18 & 1.16 \\
SMA$[\,\arcsec]$ & 3.3 & 4.0 & 4.9 \\
P.A.$[\,\degr]$ & $-64\pm7$ & $-62\pm3$ & $-50\pm 2$\\
Ecc. & $0.68\pm0.06$ & $0.71\pm0.03$ & $0.74\pm0.02$\\ 
\hline\hline
Date & \multicolumn{3}{c}{2007--03--14T00:26--00:36}\\
Filters & $V$ & $R$ & $I$ \\
\hline
Mag. &  $>23.3\pm0.2$ &  $20.8\pm0.2$ & $19.3\pm0.2$  \\
$Z_{\mathrm{p}}$ & $-0.50\pm0.01$ & $-0.39\pm0.01$ & $0.47\pm0.02$ \\
$AM$ & 1.14 & 1.14 & 1.13 \\
$ext$ & 0.13 & 0.09 & 0.05 \\
\hline
\end{tabular}
\begin{list}{}{}
\item[$^\diamond$] Total NIR magnitudes were extracted within ellipses centred on the source and defined with the semi-major axis (SMA), the position angle (P.A., relative to west axis and counter clockwise), and the eccentricity (Ecc.).
\item[$^\dagger$] The NIR extinction parameters were fixed to $JHK_{\mathrm{S}}=0.07,0.03,$ $0.03$ with an error of $\Delta ext/ext=0.2$.
\end{list}
\end{table}

\begin{figure}
\centering
\resizebox{0.86\hsize}{!}{\includegraphics{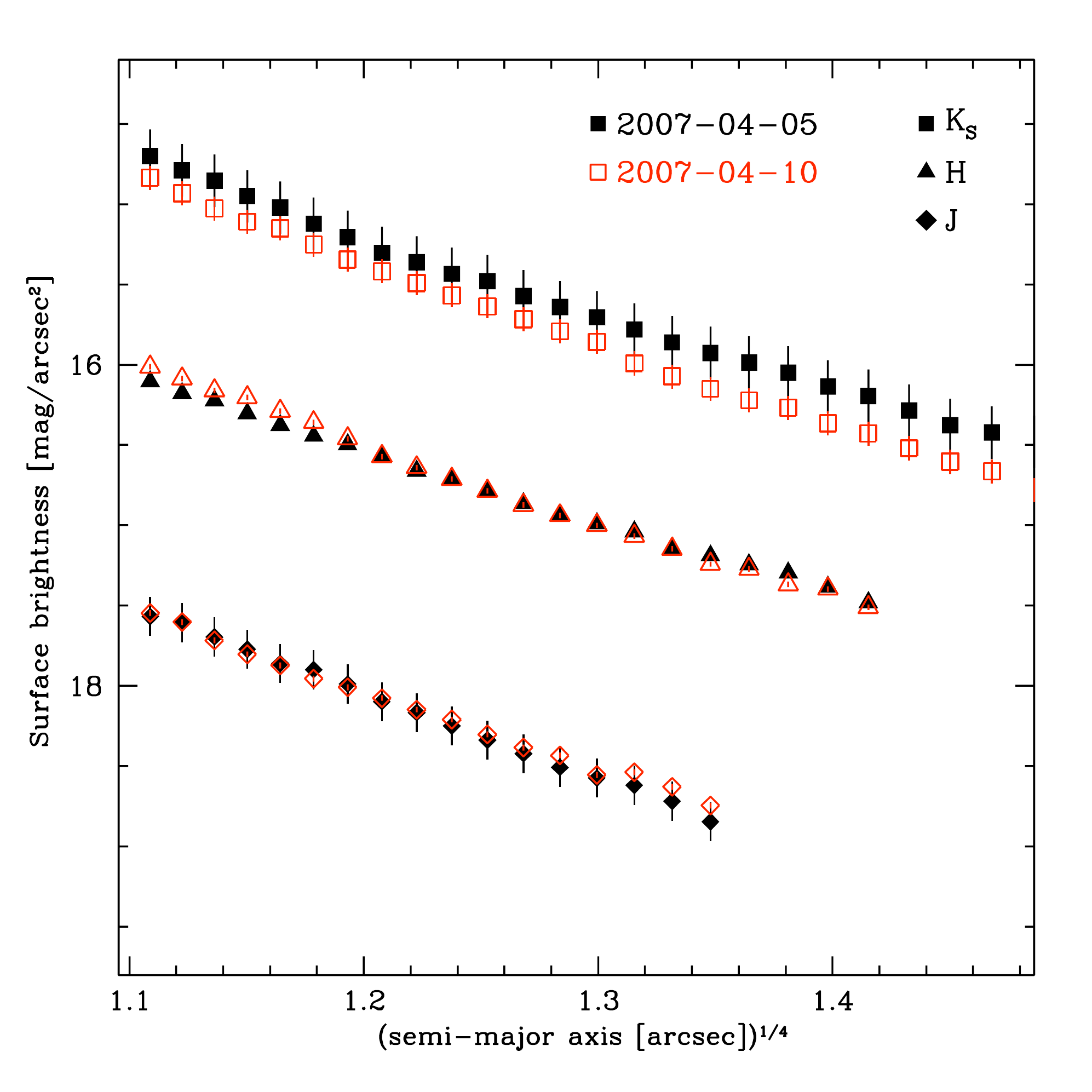}}
\caption{NIR surface brightness profiles of \igr.}
\label{brightima}
\end{figure}

\begin{table*}
\caption{NIR spectroscopy. Emission lines are represented with the laboratory $(\lambda_{0})$ and fitted $(\lambda_{\mathrm{fit}})$ wavelengths.}
\label{tab_spec}      
\centering          
\begin{tabular}{c c c c c c c c c c}
\hline\hline
~ & ~& \multicolumn{4}{c}{March 5} & \multicolumn{4}{c}{March 12} \\
\hline       
Identification & $\lambda_{0}$ & $\lambda_{\mathrm{fit}}$ & Flux & EW & FWHM & $\lambda_{\mathrm{fit}}$ & Flux & EW & FWHM\\
~ & $\mu$m & $\mu$m & arbitrary & \AA & \AA & $\mu$m & arbitrary & \AA & \AA\\
\hline
[S III] & 0.9531 & -- & -- & -- & -- & 0.9911(1) & $2724\pm226$ & $-23\pm2$ & $35\pm3$\\
He I/Pa $\gamma$ &1.0830/1.0935 & 1.1292(6) & $1418\pm96$ & $-168\pm11$ & $222\pm18$ & 1.1293(2) & $19513\pm341$ & $-155\pm3$ & $182\pm4$\\
Pa $\beta$ & 1.2820 & -- & -- & -- & -- & 1.3330(5) & $8621\pm531$ & $-59\pm4$ & $183\pm15$\\
Pa $\alpha$ & 1.8751 & 1.9466(8) & $1120\pm56$ & $-170\pm8$ & $279\pm21$ & 1.9473(2) & $12993\pm167$ & $-120\pm1$ & $269\pm4$\\
\hline
\end{tabular}
\end{table*}

The high-energy properties of the source and the NIR photometry of the 2MASS source led to a tentative classification of the source as an HMXB with a main sequence star \citep{Tomsickal08a}. However, this hypothesis is discarded by the NIR images as \igr\ appears as a non-circular extended source with an extension in the direction north-east/south-west with a semi-major axis of $4.9\arcsec$ (in $K_{\mathrm{S}}$).
The source was detected from filter $R$ to filter \Ks. The source appears point-like in the optical images. Since the X-ray observation was performed with \chandra\ that has an angular resolution similar to the optical/NIR images, it implies that we observed two different emitting components of the system between the X-rays/optical and the NIR. The most obvious candidate is thus a galaxy with an active galactic nucleus (AGN), plausibly a Seyfert galaxy, where the X-ray/optical radiation mainly comes from the accreting gas surrounding the supermassive black hole and the NIR radiation is mainly due to the dust present in the host galaxy \citep{Wilkes04}. The NIR also suggests the presence of the spiral arms of the galaxy. The NIR magnitudes (reported in Table~\ref{tab_photom}) do not show any variation between the two observations separated by 5 days, but indicate a strong reddening of the source with a NIR color $J-K=1.97\pm0.08\ (1.93\pm0.13)$~mag for the March 10th (March 5th) observation, after correcting for the Galactic extinction \citep[$E(\mathrm{B}-\mathrm{V})=1.509$,][]{Schlegelal98} and with the transformation $K-K_{\mathrm{s}}=-0.005\times(J-K)$ (p.7, SOFI manual). Such strong reddening is mainly observed in Seyfert 1 galaxies \citep{Rudyal82} and was interpreted as the reprocessing of hot dust located in the broad-line region \citep{Barvainis87}. Mid IR observations are necessary to further study the dusty component of the Seyfert galaxy. The NIR surface brightness profiles are in good agreement with the de Vaucouleurs' law ($S_{\nu}\propto R^{1/4}$, see Fig.~\ref{brightima}).

\begin{figure}
\centering
\resizebox{0.93\hsize}{!}{\includegraphics{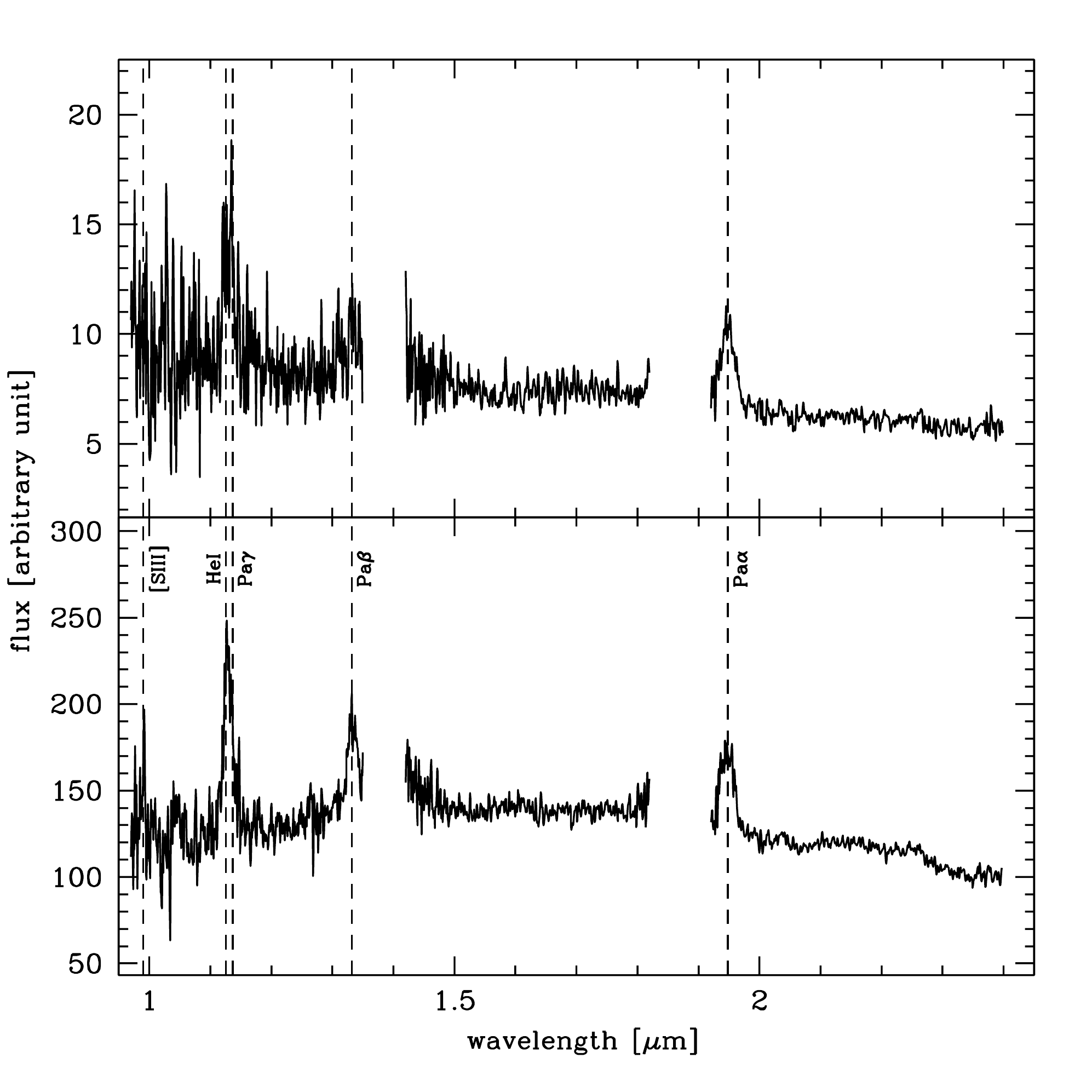}}
\caption{NIR spectra of \igr\ on March 5 ({\it top}) and March 12 ({\it bottom}), 2007. The dashed lines represent the identified emission lines red-shifted with $z=0.0391\pm0.0004$.}
\label{SPEima}
\end{figure}

The emission lines observed in the two spectra (see Fig.~\ref{SPEima}, the \emph{bottom} spectrum having a higher signal-to-noise ratio) are reported in Table~\ref{tab_spec}. The three main emission lines of the spectra situated at $\lambda\,1.1288\,\mum$, $\lambda\,1.3323\,\mum$ and $\lambda\,1.9490\,\mum$ were easily identified with the HeI $\lambda\,1.0830\,\mum$ and HI (Pa$\beta$ and Pa$\alpha$) emission lines, once red-shifted by $\sim0.4$. These are common emission lines in AGN, independent of the class \citep[see the spectral atlas of AGN in the energy range 0.8--2.4~$\mum$ of][which covers the spectral range of SOFI]{Riffelal06}. These emission lines are broad with values $>4000$~km$\,\unit{s}{-1}$. Such broad emission lines are typical of type 1 AGN.

We also tried to identify other emission lines, particularly in the spectrum taken on March 12, 2007 (see Fig.~\ref{SPEima} \emph{bottom}): 1) a narrow forbidden line [S III] $\lambda\,0.9531\,\mum$ with $FWHM=1102\pm94$~km$\,\unit{s}{-1}$, and 2) an HI Pa $\gamma$ line might be blended within the HeI $\lambda\,1.0830\,\mum$ line, still the HeI line is always the strongest one in all AGN spectra \citep[e.g. Fig.~9--12 in][]{Riffelal06}. Using only the HI Pa$\beta$ and Pa$\alpha$ emission lines and both spectra, we derived an average redshift of $z=0.0391\pm0.0004$. The line ratio of Pa$\alpha$/Pa$\beta$ gave $F(\mathrm{Pa}\alpha)/F(\mathrm{Pa}\beta)=1.5\pm0.1$. This value is slightly lower than the expected value of $1.8\pm0.1$ for Case B recombination when considering a typical broad-line cloud with temperatures between $(0.5-3)\times10^{4}$~K and electron densities between $10^{8-10}\ \unit{cm}{-3}$. Still, this is a common feature in Seyfert galaxies \citep[][and references therein]{Glikmanal06}.

\section{Conclusion}\label{secCon}
Therefore, we conclude that \igr\ is a type 1 Seyfert galaxy as demonstrated with the optical and NIR observations. The type 1 classification is in good agreement with the lack of strong absorption observed in the X-rays. Its redshift is $z=0.0391\pm0.0004$. \citet{Butleral09} reported that the distribution of redshifts for the AGNs detected by \integral\ picked at $z=0.033$ ($0.035$ for IGRs). In the current census of sources detected by \integral/ISGRI, there are 165 detected Seyfert galaxies (of which 63 are IGR sources). Discarding sources without reported redshift, the average redshift is 0.04 (0.06 when considering only IGRs) with a variance of 0.005 (0.006 for IGR) \citep{Beckmannal09}. \igr\ is in good agreement with previous IGR sources identified as Seyfert galaxies and more generally with the Seyfert galaxies detected by \integral.

\begin{acknowledgements}
JAZH thanks Farid Rahoui, Volker Beckmann and Simona Soldi for useful discussions on \igr. JAZH acknowledges the Swiss National Science Foundation for financial support. JAT acknowledges partial support from Chandra award GO8$-$9055X issued by the Chandra X-ray Observatory Center, which is operated by the Smithsonian Astrophysical Observatory for and on behalf of NASA under contract NAS8$-$03060. This work was supported by the Centre National d'Etudes Spatiales (CNES). It is based on observations with IBIS embarked on \integral. The authors acknowledge the use of 1) NASA's Astrophysics Data System, 2) the SIMBAD database, operated at CDS, Strasbourg, France, and 3) data products from the Two Micron All Sky Survey, which is a joint project of the University of Massachusetts and the Infrared Processing and Analysis Center/California Institute of Technology, funded by the National Aeronautics and Space Administration and the National Science Foundation.
\end{acknowledgements}

\bibliographystyle{aa}
\bibliography{igrj09026_arx.bbl}

\end{document}